\documentclass[a4paper]{article}

\usepackage{graphicx}

\usepackage{xr-hyper}
\usepackage[colorlinks=false]{hyperref}

\usepackage[margin=1.125in]{geometry}

 \usepackage{amsmath,amssymb,amsfonts,upref,endnotes}
\usepackage{color}

\usepackage{stackengine}

\usepackage[square,numbers]{natbib}

\title{Travelling waves in a free boundary mechanobiological\\ model of an epithelial tissue \date{}}

\author{
\footnote{Corresponding author: ryanjohn.murphy@hdr.qut.edu.au} R. J. Murphy$^{1}$, P. R. Buenzli$^{1}$, R. E. Baker$^{2}$ and  M. J. Simpson$^{1}$}

\makeatletter
\newcommand*{\addFileDependency}[1]{
  \typeout{(#1)}
  \@addtofilelist{#1}
  \IfFileExists{#1}{}{\typeout{No file #1.}}
}
\makeatother

\begin{document}

\maketitle

\vspace{-0.25in}
\begin{center}
\textit{$^{1}$ Mathematical Sciences, Queensland University of Technology, Brisbane, Australia \\
$^{2}$ Mathematical Institute, University of Oxford, Oxford, UK}
\end{center}

\begin{abstract}
 \noindent We consider a free boundary model of epithelial cell migration with logistic growth and nonlinear diffusion induced by mechanical interactions. Using numerical simulations, phase plane and perturbation analysis, we find and analyse travelling wave solutions with negative, zero, and positive wavespeeds. Unlike classical travelling wave solutions of reaction-diffusion equations, the travelling wave solutions that we explore have a well-defined front and are not associated with a heteroclinic orbit in the phase plane. We find leading order expressions for both the wavespeed and the density at the free boundary. Interestingly, whether the travelling wave solution invades or retreats depends only on whether the carrying capacity density corresponds to cells being in compression or extension. 
\end{abstract}


\newpage
\section{Introduction}
Nonlinear reaction-diffusion equations describing the dynamics of a single species often support travelling wave solutions \cite{MurrayMathBiol2002,Sherratt1990}. The classical example is the Fisher-KPP equation which has linear diffusion and a logistic reaction term \cite{Fisher1937,Maini2004a}. Travelling wave solutions of the Fisher-KPP equation are associated with heteroclinic orbits in the phase plane and correspond to invasion with a positive minimum non-dimensional wavespeed $c \geq 2$ \cite{MurrayMathBiol2002, Hilhorst2016}. Since the cell density $q(x,t) \to 0$ as $x \to \infty$, these solutions do not have compact support and do not allow us to identify a well-defined front often observed in cell invasion experiments and ecological invasion \cite{Maini2004a,Maini2004b,Kot2001,Skellam1951}.

One way of overcoming the lack of a well-defined front is to incorporate degenerate nonlinear diffusion, as in the Porous-Fisher equation \cite{Witelski1995,Sanchez-Garduno1994,Sanchez-Garduno1995,Sherratt1996}. An alternative approach to obtain travelling wave solutions with a well-defined front is to re-formulate the Fisher-KPP and Porous-Fisher models as moving boundary problems with a Stefan condition at the moving boundary \cite{Fadai2020,ElHachem2019, ElHachem2020,Fadai2020b}. Interestingly the Fisher-KPP, Porous-Fisher, and Fisher-Stefan models always lead to invading travelling waves where previously vacant regions are eventually colonised. None of these single-species models lead to retreating travelling waves where colonised regions eventually become uncolonised. Similar invading behaviour has been observed in discrete space and velocity jump processes and their continuum approximations \cite{Markham2014,Lui2010}. Retreating and invading waves have previously been observed for multi-species models \cite{Kimmel2019}.

In this work we consider a single-species model which leads to travelling wave solutions with a well-defined front that can either invade or retreat. Our free boundary model, which we derived previously \cite{Murray2009FromDimension,Baker2018ADynamics,Murphy2020a}, is motivated from a discrete model of a one-dimensional chain of epithelial cells. In this model cells are treated as mechanical springs that can be stretched or compressed and relax to a natural resting length. Cells are also able to proliferate logistically up to a maximum carrying capacity density \cite{Murray2009FromDimension,Baker2018ADynamics,Murphy2020a}. We find travelling wave solutions that are very different to the classical travelling waves of the Fisher-KPP, Porous-Fisher, or Fisher-Stefan models. We find travelling wave solutions for $-\infty < c < \infty$ which depend on the two dimensionless parameters. In the phase plane these travelling waves are not associated with heteroclinic orbits. Instead, they are associated with an orbit that leaves a saddle equilibrium node until the trajectory passes through a special point in the phase plane determined by the free boundary conditions. We find and validate analytical expressions for both the wavespeed and the density at the free boundary. Interestingly, the distinction between whether the population retreats ($c<0$) or invades ($c>0$) depends only on whether the carrying capacity density corresponds to cells being in compression or extension.  

\section{Mathematical model}
We consider a one-dimensional chain of cells forming an epithelial sheet of total length $L(t)$. Each cell can be thought to act like a mechanical spring which mechanically relaxes towards its resting cell length, $a$, according to Hooke's law. Each cell can proliferate or die logistically. Our previous work \cite{Baker2018ADynamics,Murphy2020a} shows this results in a moving boundary problem with nonlinear diffusivity, a logistic reaction term, and no-flux mechanical relaxation boundary conditions. After nondimensionalisation, the cell density, $q(x,t)>0$, which depends on position $x$ and time $t$, is governed by \cite{Baker2018ADynamics,Murphy2019,Murphy2020a}
\begin{alignat}{2}
          \frac{\partial q(x,t)}{\partial t} &= \frac{\partial}{\partial x} \left( \frac{1}{q(x,t)^2} \frac{\partial q(x,t)}{\partial x}\right) + q(x,t)\left(1 - q(x,t) \right), \quad &&0 < x < L(t), \label{eqn:dens_nondim}\\
    \frac{\partial q(x,t)}{\partial x} &= 0, \quad &&x=0, \label{eqn:dens_nondim_x0}\\
    \frac{\partial q(x,t)}{\partial x} &= \frac{q(x,t)^{3}}{\phi}\left(\frac{1}{q(x,t)} - \kappa\right) , \quad &&x=L(t), \label{eqn:dens_nondim_xl}\\
    \frac{\mathrm{d}L(t)}{\mathrm{d}t} &= -\frac{1}{q(x,t)^{3}}\frac{\partial q(x,t)}{\partial x},\quad &&x=L(t),  \label{eqn:dens_nondim_Levol}
\end{alignat}
with two dimensionless parameters $\kappa$ and $\phi$ occurring only in the free boundary condition at $x=L(t)$ in Eq.~(\ref{eqn:dens_nondim_xl}). The first, $\kappa=Ka$, is the product of the carrying capacity density, $K$, and the resting cell length, $a$, and determines whether the carrying capacity density corresponds to cells being in compression ($\kappa <1$), at the resting length ($\kappa = 1$), or in extension ($\kappa > 1$). The second, $\phi = \sqrt{\beta\eta/(4k)}$, is the ratio of the proliferation rate, $\beta$, and mechanical relaxation rate, that depends on the cell stiffness $k$ and mobility coefficient $\eta$. Eq.~(\ref{eqn:dens_nondim_Levol}) governs the evolution of the free boundary due to mechanical relaxation and mass conservation but can be thought of as a nonlinear analogue of a Stefan condition \cite{ElHachem2019,ElHachem2020,Fadai2020}. Eqs.~(\ref{eqn:dens_nondim})--(\ref{eqn:dens_nondim_Levol}) can be solved numerically by using a boundary fixing transformation \cite{Kutluay1997}, discretising the subsequent equations on a uniform mesh using a central difference approximation. The resulting system of ordinary differential equations are solved using an implicit Euler approximation, leading to a system of nonlinear algebraic equations that are solved using Newton-Raphson iteration. Key code and algorithms are available on  \href{https://github.com/ryanmurphy42/Murphy2020c.git}{GitHub}.

\section{Travelling waves}
\begin{figure}[t!]
    \centering
    \includegraphics[width=\linewidth]{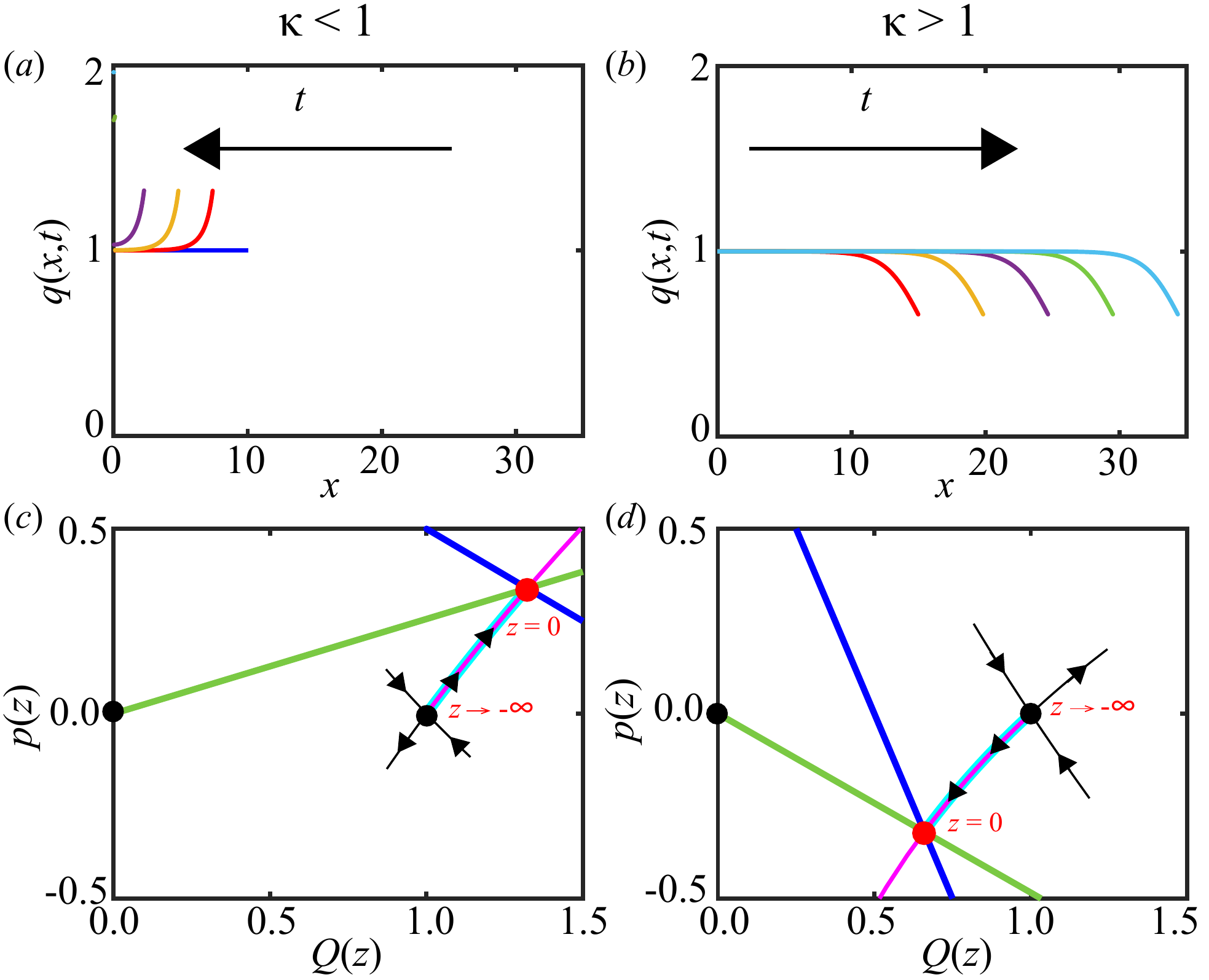}
  \caption{Travelling waves depend on $\kappa$: $c<0$ for $\kappa = 0.5 < 1$, and $c>0$ for $\kappa = 2 > 1$. (\textit{a})-(\textit{b}) Density snapshots for varying $\kappa$ at $t=0$ (blue), $10$ (red), $20$ (yellow), $30$ (purple), $40$ (green), $50$ (cyan). (\textit{c})-(\textit{d}) $(Q,p)$ phase planes for varying $\kappa$. The travelling wave solution corresponds to a trajectory governed by Eqs.~(\ref{eqn:pqfirstorder}) (magenta) between the saddle node at $(Q^{\ast},p^{\ast})=(1,0)$ from Eq.~(\ref{eqn:p_bcx0}) (black circle) and terminating at the intersection of Eq.~(\ref{eqn:p_bcxL}) (blue) and (\ref{eqn:p_evol_xL}) (green) given by Eq.~(\ref{eqn:terminationpoint}) (red circle). Continuum solution from Eqs.~(\ref{eqn:dens_nondim})--(\ref{eqn:dens_nondim_Levol}) (cyan line). The degenerate node $(Q^{\ast},p^{\ast})=(0,0)$ is also shown (black circle). All results for $\phi=1$.}
    \label{fig:Fig1}
\end{figure}

In Figure \ref{fig:Fig1} we present numerical solutions of Eqs.~(\ref{eqn:dens_nondim})--(\ref{eqn:dens_nondim_Levol}) for varying $\kappa$ and initial density condition $q(x,0) = 1$ for $0 < x < L(0) = 10$, which remains uniform and stationary, with $c=0$ for $t > 0$, when $\kappa=1$. The numerical results in Figure 2(\textit{a}) suggest the emergence of travelling wave solutions with $c<0$ when $\kappa < 1$ (Figure \ref{fig:Fig1}(a)), with $c=0$ when $\kappa = 1$ (not shown), and $c>0$ when $\kappa > 1$ (Figure \ref{fig:Fig1}(b)). The travelling waves form after initial transient behaviour. For $\kappa>1$ the invading travelling waves in the numerical simulations continue as $t \to \infty$. For $\kappa<1$ we observe retreating travelling wave-like behaviour with $c<0$ for some intermediate time before $L(t)$ approaches $x=0$ and boundary effects play a role (not shown). The cell density at the free boundary is $Q_{L} > 0$.

After a travelling wave has formed $L(t)\sim ct$ (Figure \ref{fig:FigS6}), where $c$ is the constant speed of propagation, and we introduce travelling wave coordinates $z = x - ct$. Letting $Q(z)=q(x,t)$ then Eq.~(\ref{eqn:dens_nondim}) becomes
\begin{equation}\label{eqn:dens_twc}
    \begin{split}
        \frac{\mathrm{d}}{\mathrm{d}z}\left( \frac{1}{Q(z)^2} \frac{\mathrm{d}Q(z)}{\mathrm{d}z}\right) + c \frac{\mathrm{d}Q(z)}{\mathrm{d}z} + Q(z)\left( 1 - Q\left(z\right)\right) = 0, \quad -\infty < z < 0.
    \end{split}
\end{equation}
where we choose $z=0$ to correspond to the free boundary at $x=L(t)$.

To analyse Eq.~(\ref{eqn:dens_twc}) in the two dimensional phase plane we let $p(z) = (1/Q(z)^2) \ \mathrm{d}Q(z)/\mathrm{d}z$ \cite{Li2020} to give
\begin{align}\label{eqn:pqfirstorder}   
       \frac{\mathrm{d}Q}{\mathrm{d}z} = p Q^2,   \qquad  \frac{\mathrm{d}p}{\mathrm{d}z} = Q\left[ -cpQ - \left(1 - Q \right)\right].
\end{align}
The dynamical system given by Eqs.~(\ref{eqn:pqfirstorder}) has two equilibrium points. The first at $(Q^{\ast},p^{\ast})=(0,0)$ is a degenerate node. The second at $(Q^{\ast},p^{\ast})=(1,0)$ is a saddle node for $c\neq 0$ and a degenerate node when $c=0$. Interestingly, in contrast to the Fisher-KPP equation \cite{MurrayMathBiol2002}, here linear stability analysis provides no restrictions on $c$.

We return to the boundary conditions from Eqs.~(\ref{eqn:dens_nondim_x0})--(\ref{eqn:dens_nondim_Levol}), and after transforming to travelling wave coordinates and writing in terms of $p$, we obtain
\begin{alignat}{2}
    (Q,p) &= (1,0), \quad &&z\to-\infty, \label{eqn:p_bcx0}\\
        p &= \frac{1}{\phi}\left(  1 - \kappa Q \right), \quad &&z=0, \label{eqn:p_bcxL}\\
        p &= - cQ, \quad &&z=0,\label{eqn:p_evol_xL}
\end{alignat}
where Eq.~(\ref{eqn:p_bcx0}) is informed by numerical travelling wave solutions in Figure \ref{fig:Fig1}.

In Figures \ref{fig:Fig1}(c),(d) we generate the $(Q,p)$ phase plane for $\kappa <1$ and $\kappa > 1$, respectively, using MATLAB functions quiver and ode45 \cite{MATLAB}. Trajectories corresponding to travelling wave solutions are initiated on the relevant eigenvector associated with the saddle node. We find that travelling wave solutions correspond to phase plane trajectories that run between $(Q^{\ast},p^{\ast})=(1,0)$, and a special point given by the intersection of Eqs.~(\ref{eqn:p_bcxL}) and (\ref{eqn:p_evol_xL}) given by 
\begin{equation}\label{eqn:terminationpoint}
    \left(Q_{L},p_{L}\right) = \left( \frac{1}{\kappa - c\phi}, \frac{-c}{\kappa - c\phi}\right).
\end{equation}
The remainder of the trajectory beyond $\left(Q_{L},p_{L}\right)$, obtained by solving Eqs. (\ref{eqn:pqfirstorder}), corresponds to $z>0$ and is not associated with the travelling wave solution which is restricted to $z \leq 0$. The part of the trajectory with $z>0$ tends to infinity rather than to the degenerate equilibrium point at $(Q^{\ast},p^{\ast})=(0,0)$. Therefore, the travelling wave solution is not associated with a heteroclinic orbit. This is very interesting as classical travelling waves solutions are associated with heteroclinic orbits in the phase plane \cite{MurrayMathBiol2002}-\cite{Fadai2020b}.

\begin{figure}
    \centering
    \includegraphics[width=\linewidth]{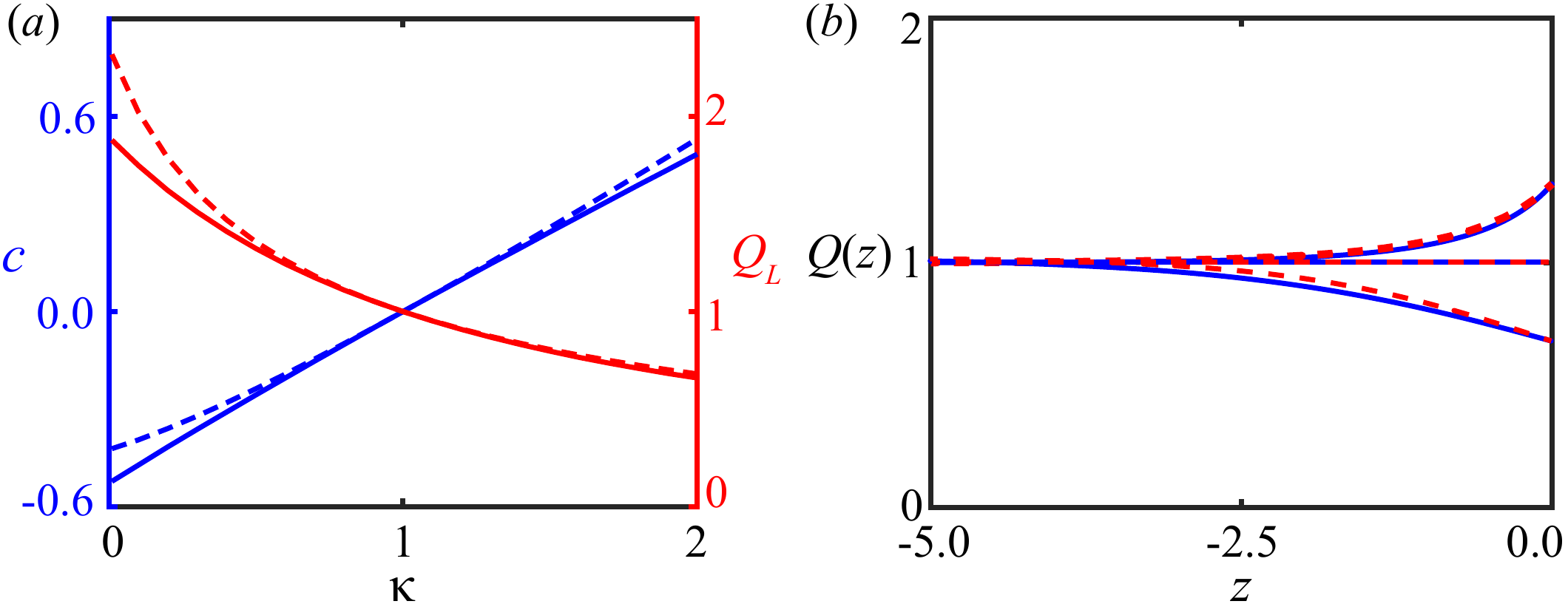}
  \caption{Travelling wave perturbation analysis. (\textit{a}) Properties of the travelling wave. Wavespeed $c$ as a function of $\kappa$ (blue) and density at free boundary $Q_{L}$ as a function of $\kappa$ (red).  Solid lines: continuum model given by Eqs.~(\ref{eqn:dens_nondim})--(\ref{eqn:dens_nondim_Levol}). Dashed lines: leading order implicit solution given by Eq.~(\ref{eqn:leadingorderc_exact}). (\textit{b}) Travelling wave solutions for $\kappa=0.5$ (top), $\kappa=1$ (middle), $\kappa=2$ (bottom) obtained by continuum model from Eqs.~(\ref{eqn:dens_nondim})--(\ref{eqn:dens_nondim_Levol}) (blue solid) and leading-order perturbation solution from {Eq.~(\ref{eqn:p0Q})} (red dashed). All for fixed $\phi = 1$.}
    \label{fig:Fig2}
\end{figure}

To provide insight into the travelling wave solutions in Figure \ref{fig:Fig1} we now seek to determine a relationship between $c$, $\kappa$, and $\phi$. By solving the continuum model we expect $c \to 0$ as $\kappa \to 1$ (Figure \ref{fig:Fig2}(\textit{a})). Therefore, we seek a perturbation solution $ p(Q) = p_{0}(Q) + c p_{1}(Q) + \mathcal{O}\left(c^2 \right)$ for  $|c| \ll 1$ which we substitute into the equation for $\mathrm{d}p/\mathrm{d}Q$ determined from Eqs.~(\ref{eqn:pqfirstorder}) to find 
\begin{equation}\label{eqn:p0Q}
    \begin{split}
        p_{0}(Q) = \pm \sqrt{ 2\left[ Q - \log_e(Q) - 1 \right]},
    \end{split}
\end{equation}
where the positive root corresponds to $c<0$ and the negative root corresponds to $c>0$. The integration constant is chosen such that Eq.~(\ref{eqn:p0Q}) satisfies Eq.~(\ref{eqn:p_bcx0}). Eq.~(\ref{eqn:p0Q}) corresponds to a small-$c$ approximation of the unstable manifold of the saddle point $(1,0)$. Applying the free boundary condition from Eq.~(\ref{eqn:p_evol_xL})
and using $Q_{L}$ from Eq.~(\ref{eqn:terminationpoint}) gives
\begin{equation}\label{eqn:leadingorderc_exact}
    \begin{split}
       |c| = \left(\kappa - c\phi\right)\sqrt{2\left[\frac{1}{\kappa - c\phi} - \log_e\left(\frac{1}{\kappa - c\phi}\right) - 1\right]}.
    \end{split}
\end{equation}
Eq.~(\ref{eqn:leadingorderc_exact}) can be solved implicitly for $c$ as a function of $\kappa$ and $\phi$ and provides good agreement with the long time numerical solutions of Eqs.~(\ref{eqn:dens_nondim})--(\ref{eqn:dens_nondim_Levol}) (Figure \ref{fig:Fig2}). To find an approximate explicit form for $c$ and $Q_{L}$, we expand Eq.~(\ref{eqn:leadingorderc_exact}) about $\kappa - c\phi=1$, and use Eq.~(\ref{eqn:terminationpoint}) to give
\begin{equation}\label{eqn:wavespeedleadingorder}
    \begin{split}
        c = \frac{\kappa - 1}{\phi + 1} + \mathcal{O}\left( \left(\kappa - c\phi -1\right)^{3/2}\right), \qquad  Q_{L} = \frac{1 + \phi}{\kappa + \phi}.
    \end{split}
\end{equation}
We find these leading order expressions in Eq.~(\ref{eqn:wavespeedleadingorder}) to be accurate close to $\phi=1$ (Figures \ref{fig:Fig1}-\ref{fig:FigS5}).

In Figure \ref{fig:Fig2}b we plot the shape of the travelling wave obtained by considering long time numerical solutions of Eqs.~(\ref{eqn:dens_nondim})--(\ref{eqn:dens_nondim_Levol}) and compare this to the leading order perturbation solution. The leading order perturbation solution is obtained by solving Eq.~(\ref{eqn:p0Q}) with the definition of $p(z)$ in Eqs. (\ref{eqn:pqfirstorder})  together with Eq.~(\ref{eqn:terminationpoint}) as the initial condition. We observe excellent agreement for $|c| \ll 1$ about $\kappa=1$.

In summary, by considering a reaction-diffusion equation arising from a biologically motivated discrete model, we find an interesting result where whether a population invades or retreats corresponds to whether cells at the carrying capacity density are in compression or in extension, respectively. We also obtain exact expressions for the speed of travelling wave solutions of Eqs,~(\ref{eqn:dens_nondim})--(\ref{eqn:dens_nondim_Levol}), together with useful approximations of the shape of the travelling wave solutions when $|c| \ll 1$. We do not pursue an existence proof of these travelling wave solutions here, but leave this for future consideration \cite{Kot2001}.

\section*{Acknowledgements}
This work was funded by the Australian Research Council (DP200100177,DP180101797). R.E.B is a Royal Society
Wolfson Research Merit Award holder, would like to thank the Leverhulme Trust for a Research
Fellowship and also acknowledges the BBSRC for funding via grant no. BB/R000816/1. We thank the two referees for their helpful comments.

\newpage
\appendix
\section{Supplementary Figures}

\begin{figure}[h!]
    \centering
    \includegraphics[width=\linewidth]{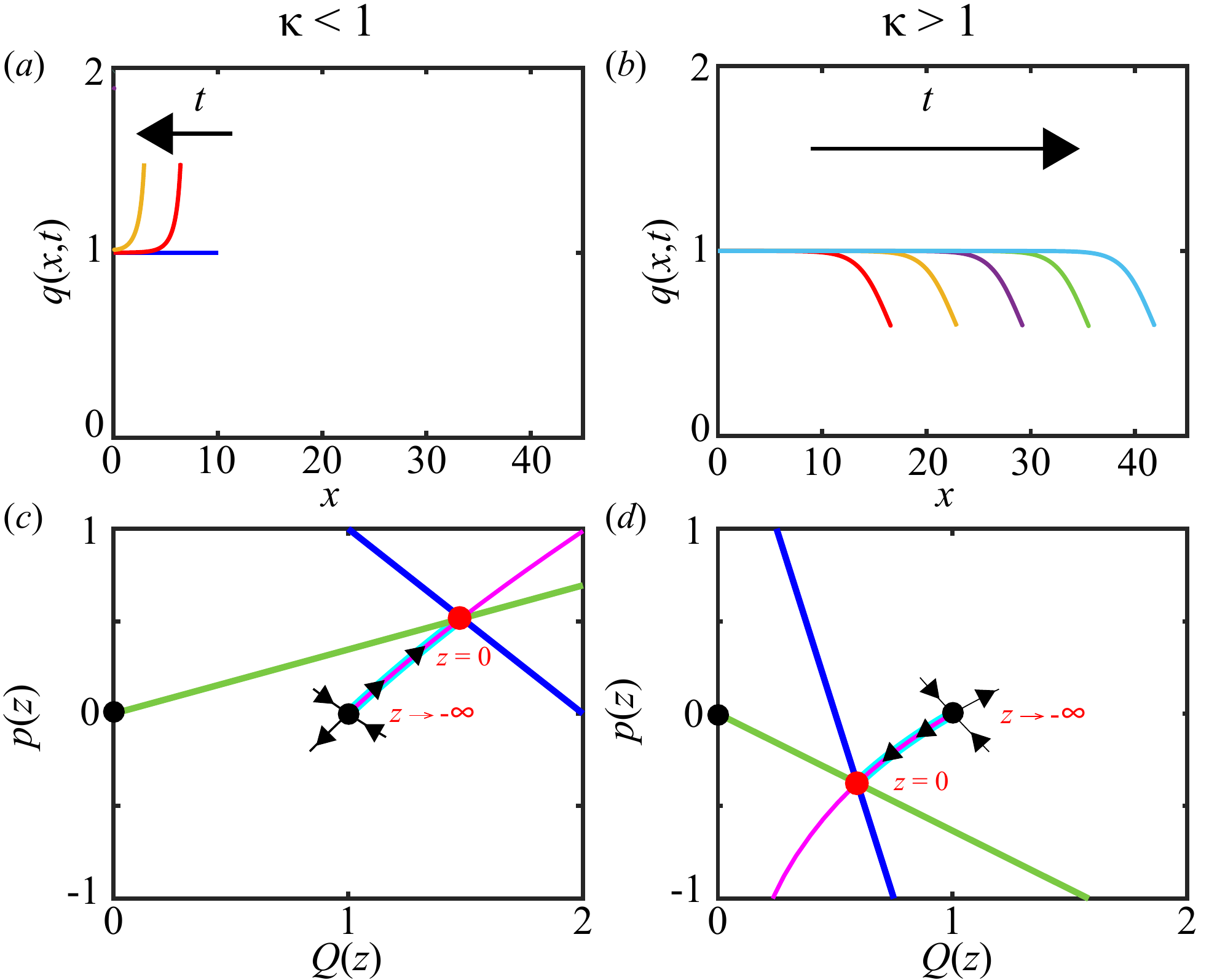}
  \caption{Results for $\phi=0.5$. Travelling waves depend on $\kappa$: $c<0$ for $\kappa = 0.5 < 1$, and $c>0$ for $\kappa = 2 > 1$. (\textit{a})-(\textit{b}) Density snapshots for varying $\kappa$ at $t=0$ (blue), $10$ (red), $20$ (yellow), $30$ (purple), $40$ (green), $50$ (cyan). (\textit{c})-(\textit{d}) $(Q,p)$ phase planes for varying $\kappa$. The travelling wave solution corresponds to a trajectory governed by Eqs.~(6) (magenta) between the saddle node at $(Q^{\ast},p^{\ast})=(1,0)$ from Eq.~(7) (black circle) and terminating at the intersection of Eq.~(8) (blue) and 9) (green) given by Eq.~(10) (red circle). Continuum solution from Eqs.~(1)--(4) (cyan line). The degenerate node $(Q^{\ast},p^{\ast})=(0,0)$ is also shown (black circle).}
    \label{fig:FigS1}
\end{figure}

\begin{figure}[h!]
    \centering
    \includegraphics[width=\linewidth]{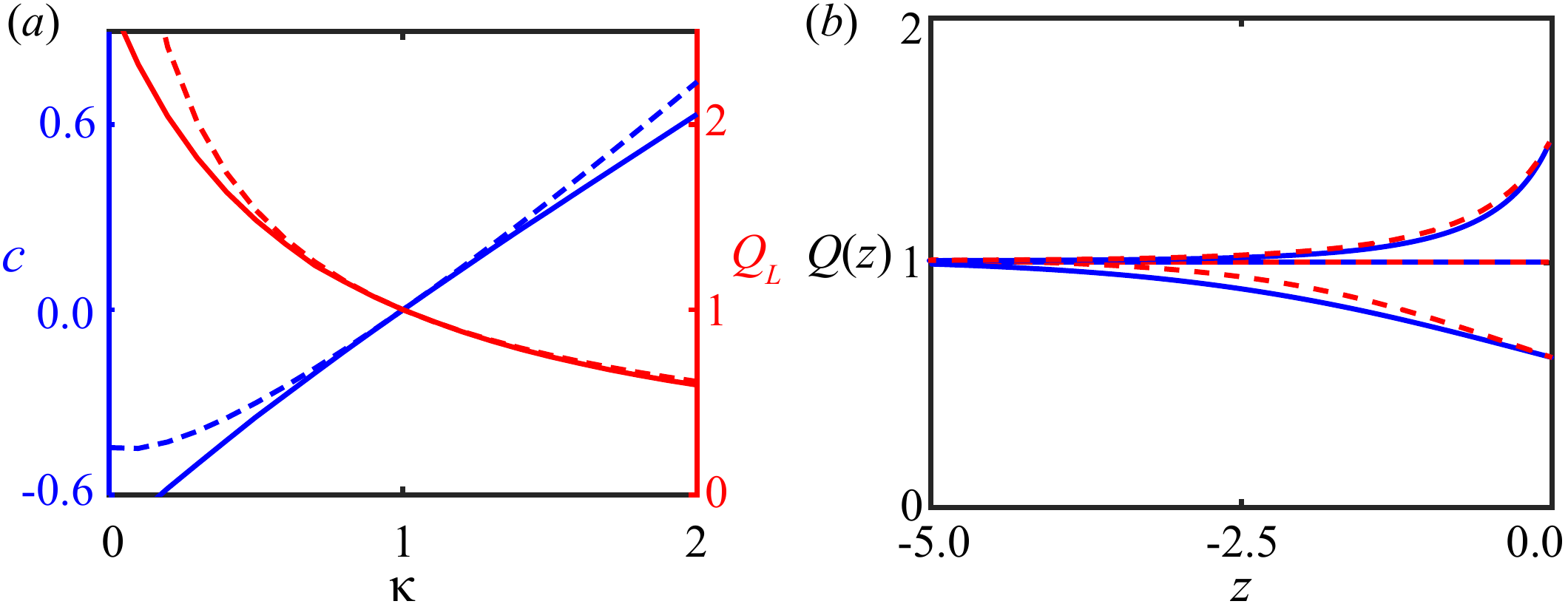}
  \caption{Travelling wave perturbation analysis  for $\phi=0.5$. (\textit{a}) Properties of the travelling wave. Wavespeed $c$ as a function of $\kappa$ (blue) and density at free boundary $Q_{L}$ as a function of $\kappa$ (red).  Solid lines: continuum model given by Eqs.~(1)--(4). Dashed lines: leading order implicit solution given by Eq.~(12). (\textit{b}) Travelling wave solutions for $\kappa=0.5$ (top), $\kappa=1$ (middle), $\kappa=2$ (bottom) obtained by continuum model from Eqs.~(1)--(4) (blue solid) and leading-order perturbation solution from {Eq.~(11)} (red dashed).}
    \label{fig:FigS2}
\end{figure}

\begin{figure}[h!]
    \centering
    \includegraphics[width=\linewidth]{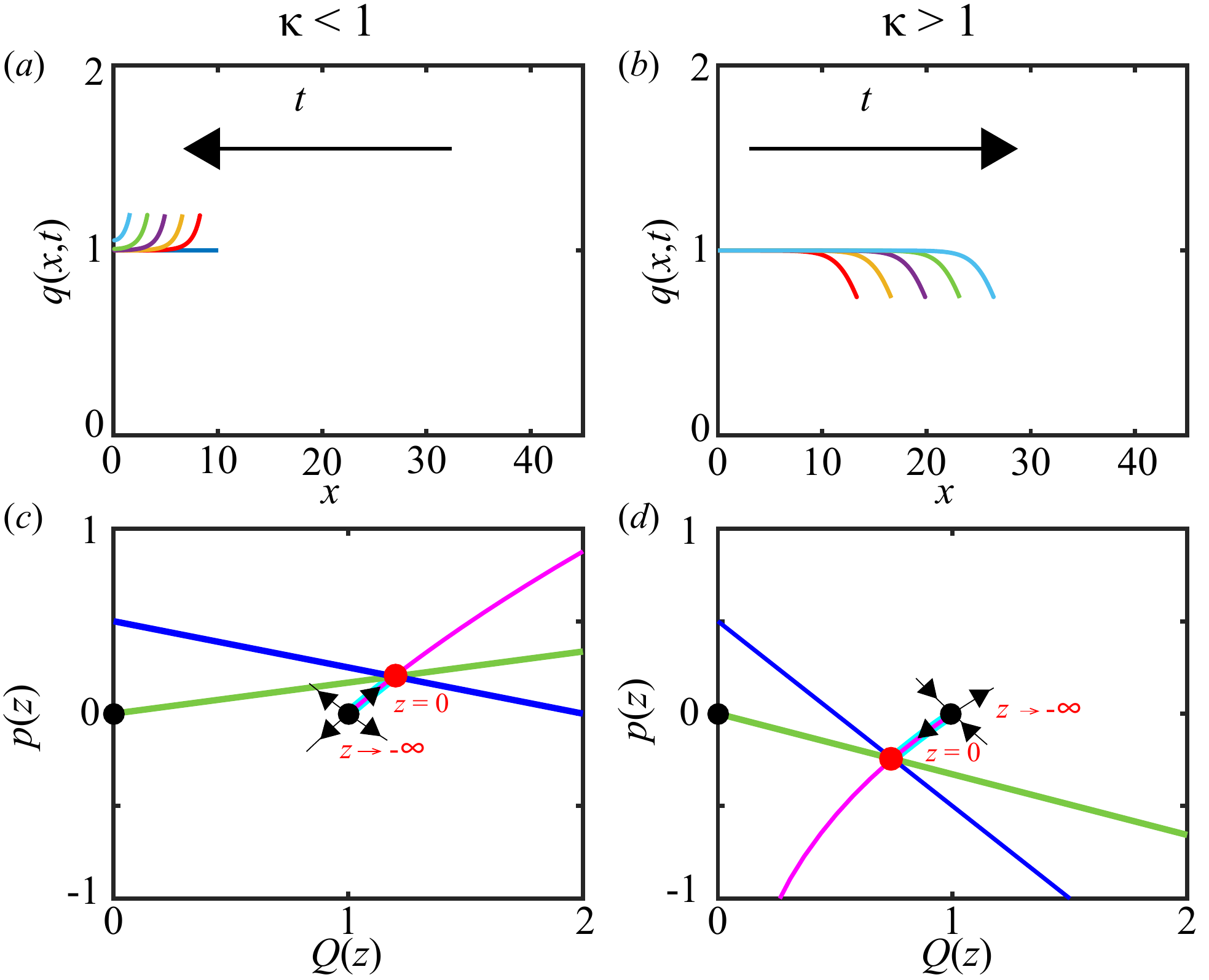}
  \caption{Results for $\phi=2$. Travelling waves depend on $\kappa$: $c<0$ for $\kappa = 0.5 < 1$, and $c>0$ for $\kappa = 2 > 1$. (\textit{a})-(\textit{b}) Density snapshots for varying $\kappa$ at $t=0$ (blue), $10$ (red), $20$ (yellow), $30$ (purple), $40$ (green), $50$ (cyan). (\textit{c})-(\textit{d}) $(Q,p)$ phase planes for varying $\kappa$. The travelling wave solution corresponds to a trajectory governed by Eqs.~(6) (magenta) between the saddle node at $(Q^{\ast},p^{\ast})=(1,0)$ from Eq.~(7) (black circle) and terminating at the intersection of Eq.~(8) (blue) and (9) (green) given by Eq.~(10) (red circle). Continuum solution from Eqs.~(1)--(4) (cyan line). The degenerate node $(Q^{\ast},p^{\ast})=(0,0)$ is also shown (black circle).}
    \label{fig:FigS3}
\end{figure}

\begin{figure}[h!]
    \centering
    \includegraphics[width=\linewidth]{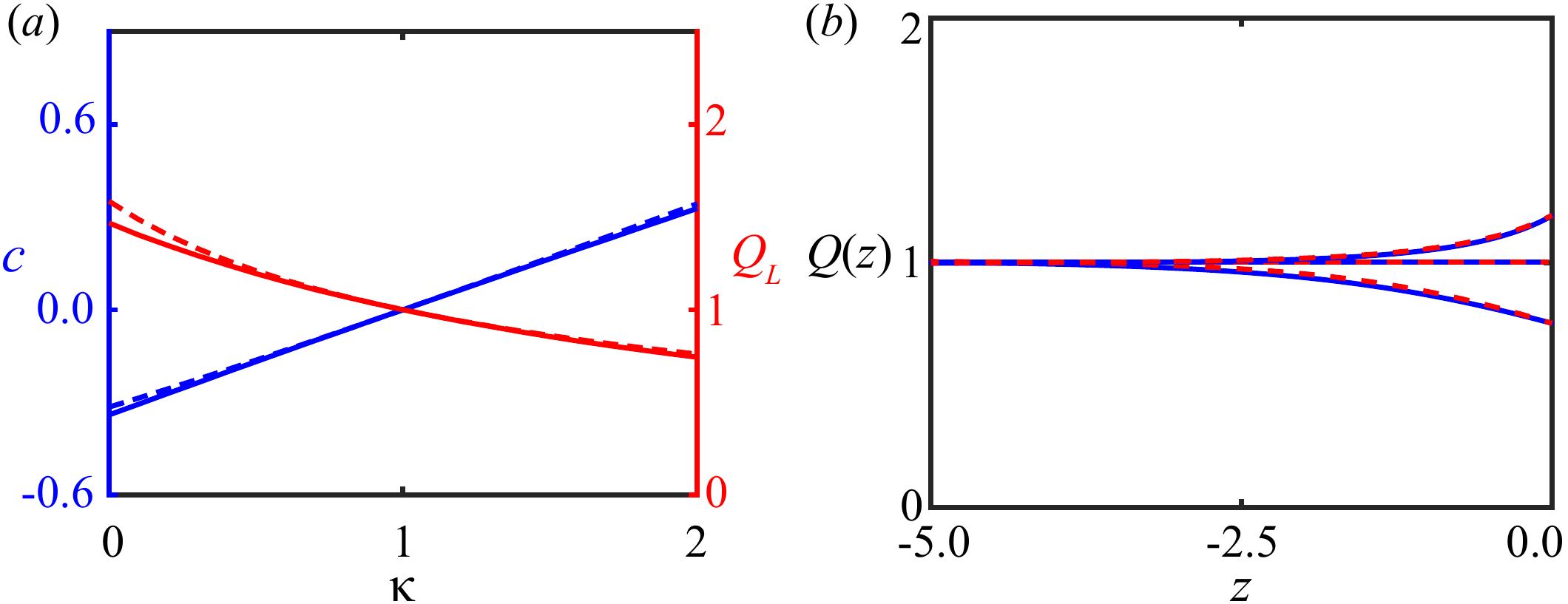}
  \caption{Travelling wave perturbation analysis for $\phi=2$. (\textit{a}) Properties of the travelling wave. Wavespeed $c$ as a function of $\kappa$ (blue) and density at free boundary $Q_{L}$ as a function of $\kappa$ (red).  Solid lines: continuum model given by Eqs.~(1)--(4). Dashed lines: leading order implicit solution given by Eq.~(12). (\textit{b}) Travelling wave solutions for $\kappa=0.5$ (top), $\kappa=1$ (middle), $\kappa=2$ (bottom) obtained by continuum model from Eqs.~(1)--(4) (blue solid) and leading-order perturbation solution from {Eq.~(11)} (red dashed). All for fixed $\phi = 2$.}
    \label{fig:FigS4}
\end{figure}

Results for Figure 1 and 2 in the manuscript are presented for $\phi=1$. We now reproduce these figures for $\phi=0.5$ and $\phi=2$ in Figures \ref{fig:FigS1}, \ref{fig:FigS2} and Figures \ref{fig:FigS3}, \ref{fig:FigS4}, respectively.

\begin{figure}[h!]
    \centering
    \includegraphics[width=\linewidth]{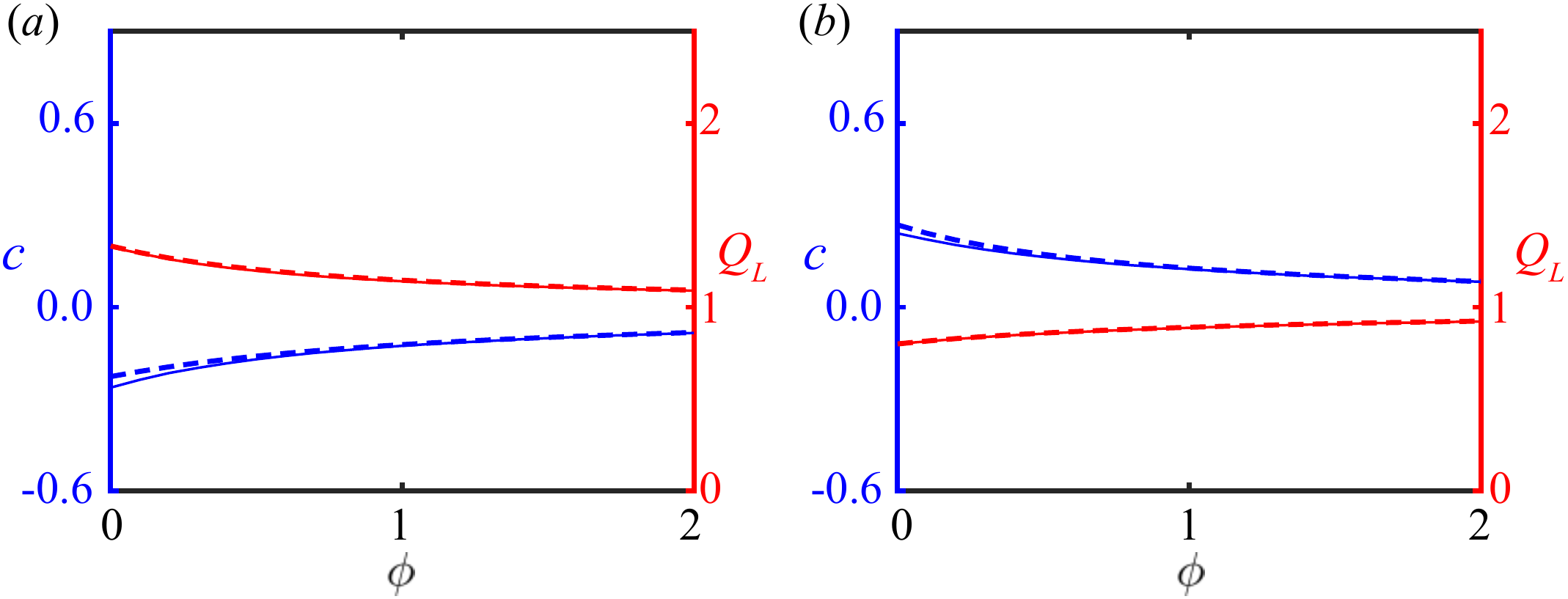}
  \caption{Travelling wave perturbation analysis. Properties of the travelling wave. Wavespeed $c$ as a function of $\phi$ (blue) and density at free boundary $Q_{L}$ as a function of $\phi$ (red) for (\textit{a}) $\kappa=0.75$ and (\textit{b}) $\kappa=1.25$. Solid lines: continuum model given by Eqs.~(1)--(4). Dashed lines: leading order implicit solution given by Eq.~(12).}
    \label{fig:FigS5}
\end{figure}

In Figure \ref{fig:FigS5} we plot the dependence of the wavespeed, $c$, and density at the boundary, $Q_{L}$, against $\phi$ for $\kappa=0.75$ and $\kappa=1.25$.

\begin{figure}[h!]
    \centering
    \includegraphics[width=\linewidth]{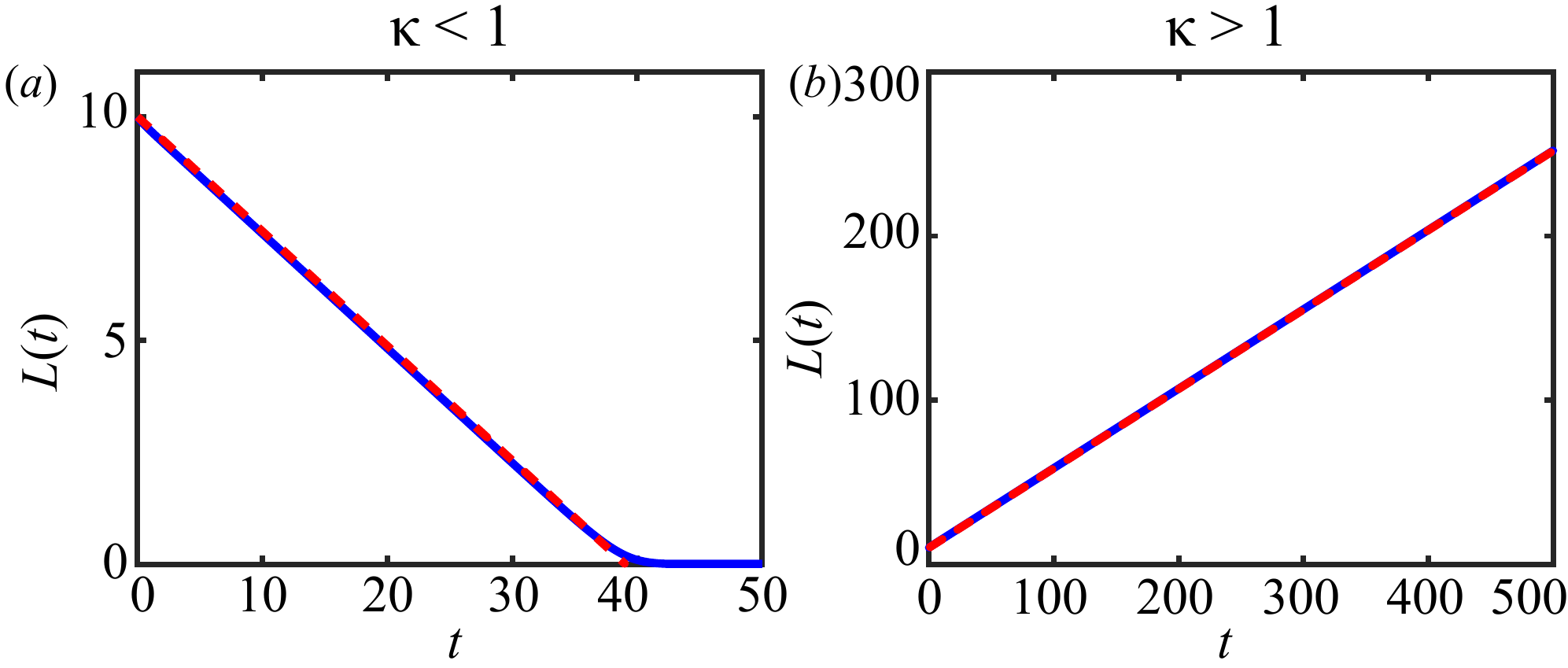}
  \caption{Evolution of tissue length $L(t)$. Comparison of continuum model given by Eqs.~(1)--(4) (solid blue lines) with $L(t) \sim ct$ (red dashed lines). (\textit{a}) $\kappa = 0.5 < 1$ and $L(t) = 10 + ct$ where $c=0.484$. (\textit{b}) $\kappa = 2 > 1$ and $L(t) = 10 + ct$ where $c=-0.256$. Solutions correspond to Figure 1 for $\phi = 1$.}
    \label{fig:FigS6}
\end{figure}

In Figure \ref{fig:FigS6}, we support the statement that after the travelling waves have formed $L \sim ct$ by plotting $L(t)$ against $t$ for the results corresponding to Figure 1. Similar excellent agreement is found for other results (not shown).

\newpage
\clearpage
\bibliographystyle{elsarticle-num-names}
\bibliography{references}

\end{document}